\documentclass[twocolumn,showpacs,pre]{revtex4} 
\usepackage{graphicx}% Include figure files
\usepackage{dcolumn}% Align table columns on decimal point
\usepackage{bm}% bold math
\begin{document} 

\title{Measurement of synchronization properties in systems with different excitation levels}

\author{Micha{\l}\ \.Zochowski and Rhonda Dzakpasu} 
\affiliation{Department of Physics and Biophysics Research Division\\
University of Michigan\\
Ann Arbor, MI 48109}

\begin{abstract}
We devised a measure based on the distributions of relative event timings of two coupled units. The measure dynamically evaluates temporal interdependencies between the two coupled units. Using this we show that even in the event of symmetrical coupling, non-identical units (having different control parameters) have small but persistent shifts in their event timings that change dramatically with the relative properties of their internal trajectories. We link that finding with the activity dependent synaptic modification in the brain showing that internally or externally driven excitation levels of the neuron may determine direction of information processing in a network.     
\end{abstract} 
\date{\today}
\maketitle 
It is hypothesized that the temporal structure of activity in the brain can play a crucial role in information processing \cite{malsburg2}. It was also established that relative timings between the spiking of coupled neurons may lead to long- and/or short- term synaptic modifications \cite {poo, mcmah, abbott01}. To be able to monitor successfully such spatio-temporal patterning in biological systems in general it is crucial to develop adequate methods to measure relative timings of the observed events. Any method designed for this purpose must be statistical, based on information obtained from many events but at the same time able to detect dynamical changes in the system. We have developed such a measure that is based on the dynamical monitoring of entropy changes in relative firing patterns between two coupled units. We will refer to it as a conditional entropy (CE) since it is calculated based on the relative timings of the events of both coupled units. 
\begin{figure}
\includegraphics[scale=1.0]{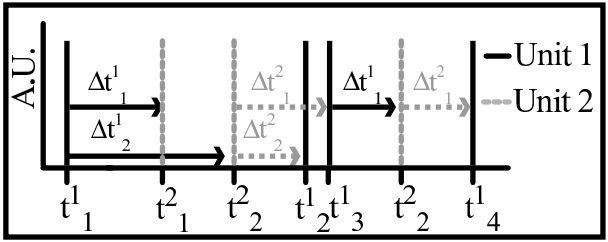}
\caption{Relative event timings used to calculate the conditional entropy. The timings of one unit (or neuron) are calculated in relation to the timing of the last event of the other one.}
\end{figure}
The relative timings measured for the two units are shown in Fig. 1. When the conditional entropy of the event timing pattern of unit 2 was calculated with respect to unit 1, we measured the timings of events that occurred at unit 2 with respect to the last event of unit 1. Conversely when the conditional entropy of unit 1 with respect to unit 2 was calculated, we measured the event timing of unit 1 with respect to the last event of unit 2. The probability distributions are dynamically updated (when the new event takes place) by increasing the bin of the distribution in which the latest timing falls by a fixed $\Delta P$. Thus, if the timing falls in the bin $I$,
$P_I(t)=P_I(t-1)+\Delta P$.
The distributions were then renormalized.  Thus, the $\Delta P$ effectively determines to what degree the distribution will be skewed towards the timings of the newest events, so that the probability of the event timing after $n$ events declines to:
\begin{equation}
P_q(t)=\frac{1}{(1+\Delta P)^n}P_q(t-n).
\label{eq2}
\end{equation}
The entropies are calculated from the normalized distributions, $S=-\sum_{I}P_I\ln{P_I}$, every time new events occur.  
We applied the measure to monitor relative timing patterns of two coupled non-identical R{\"o}ssler oscillators \cite{rossler} and then two Hindmarsh-Rose models of the neurons \cite{hindmarsh}. The coupled units differed through the value of their control parameters that influence dynamical properties of their trajectories. In light of the second example we will relate to those properties as excitation levels.

The event timings for the R{\"o}ssler oscillator were defined as the times at which the oscillator's trajectory (unit 1 or 2) crosses specified Poincare section ($z=1$). 
The equations for the coupled R{\"o}ssler system are:
\begin{equation}
\begin{array}{l}
\dot{x}_{1,2}=-(z_{1,2}+y_{1,2})\\
\dot{y}_{1,2}=x_{1,2}+a_{1,2}y_{1,2}+ \alpha (y_{2,1}-y_{1,2})\\
\dot{z}_{1,2}=b+(x_{1,2}-c)z_{1,2} 
\end{array}
\end{equation}
where the subscript denotes the oscillator number; $a_{1,2}$, $b=0.2$, $c=10.0$ are the control parameters of the oscillators and $\alpha=0.4$ is the coupling. The values of parameters $a_1$ and $a_2$ were different for both units. 
\begin{figure}
\includegraphics[scale=.55]{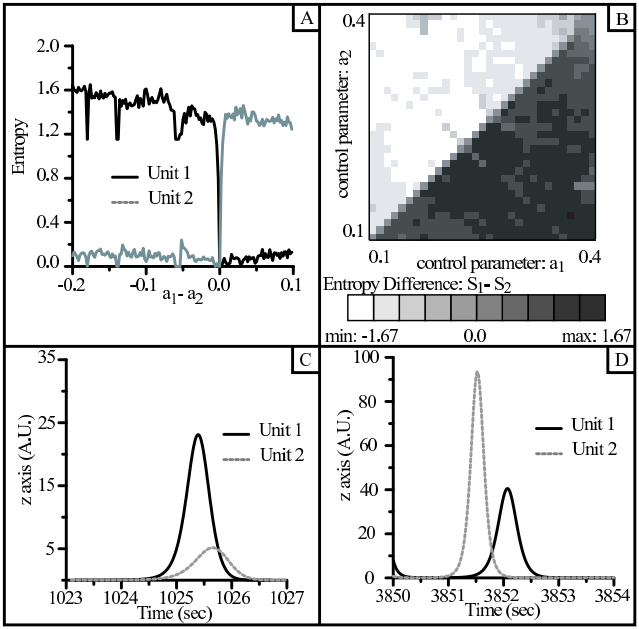}
\caption{Conditional entropy calculated for two R{\"ossler} oscillators with different relative excitation levels (value of $a_{1,2}$). Panel A: the $a_1$ is varied from 0.1 to 0.4 and the control parameter of unit 2 is fixed and $a_2=0.3$. Panel B: is a grayscale map of the conditional entropy difference when both $a_{1,2}$ are varied from 0.1 to 0.4. Panels C and D: The relative timings of the spikes for the two units when the excitation: C)$a_{1}>a_{2}$ and D) $a_{2}>a_{1}$.}
\end{figure}

We found that the measure highlighted an important result. The conditional entropy of the oscillator with higher excitation level (a larger control parameter, $a_i$) is close to zero, whereas the entropy of the oscillator with lower excitation is significantly higher (Fig 2A). Moreover, a small change in the difference between the control parameters that changes their relative excitation levels causes the conditional entropies to change rapidly. We applied the measure to coupled oscillators having parameter $a_{1,2}$ span an extended range of values (Fig 2B); the observed phenomenon persisted over whole the range. Fig 2B plots grayscale map to represent the difference between the conditional entropies ($S_1-S_2$) for the full range of the parameters of both units; white denotes minimal value whereas black denotes the maximal one. The difference values do not change significantly except in the neighborhood of the diagonal ($a_1=a_2$), where $S_1=S_2=0$. As the diagonal is crossed the values of the entropies switch and thus their difference takes opposite values. This is due to the fact that even though the coupling between the two units is symmetrical, there is asymmetry in the firing pattern that depends differentially on the relative excitation level of the coupled oscillators. That is, even if the firing seems to be coincident due to the coupling, the event timing of the unit that has higher excitation precedes narrowly that of the other unit. Fig 2C, D depicts this situation. We integrated the trajectories of the two oscillator for $5000$s; at $t=2500$ we changed the values of the control parameters so that at $t<2500$s, $a_1<a_2$ and at $t>2500$s $a_1>a_2$. We plot an example of the relative timing of the events of the two units for both regimes. Only at $a_1 = a_2$, the two units synchronize with zero time lag, making their conditional entropies equal and zero. Fig. 3 shows the changes in the conditional entropies of the two units due to changes of relative values of the control parameter $a$. As the relative values of control parameter changes, the entropies change dramatically following the changes of the timing interdependencies of the two units: $\alpha=0$ in I and V sections - the two units are independent and both have high conditional entropies; $\alpha=0.4$ and the $a_{1}=a_2$ in section II - the conditional entropies of both units decrease to zero; when the units have different relative excitation levels (sections III and IV), their conditional entropies start changing rapidly at the onset of the sections, with the conditional entropy of the unit with higher excitation level tending to zero, and that of the other one being high.
\begin{figure}
\includegraphics[scale=.76]{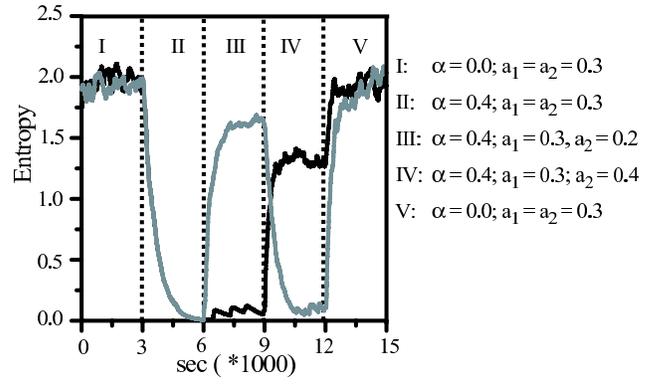}
\caption{The changes of conditional entropy are monitored during different phases of the same integration. The duration of the interaction was 16000 seconds and there were 5 intervals of 3000s  each. The settings of every interval are listed above.}
\end{figure}

We investigated the stability of the described phenomena for different values of coupling (Fig 4). As before, the control parameter of one unit was constant and that of the second unit was varied. When the coupling is equal to zero both oscillators are independent and their conditional entropies are high (Fig. 4A). However, the described timing interdependencies materialize even for small couplings.  For a coupling as low as $\alpha=0.2$ the conditional entropies for both oscillators are significantly different (Fig. 4B). The conditional entropies $S_1=S_2$ and are equal to zero only when both oscillators have the same values of their control parameters (Fig 4C). Thus, this is the only time when the two oscillators mutually synchronize. At larger couplings the events of the oscillator with higher excitation precede within the narrow window the ones of the oscillator with lower excitation (Fig. 4D).
\begin{figure}
\includegraphics[scale=.58]{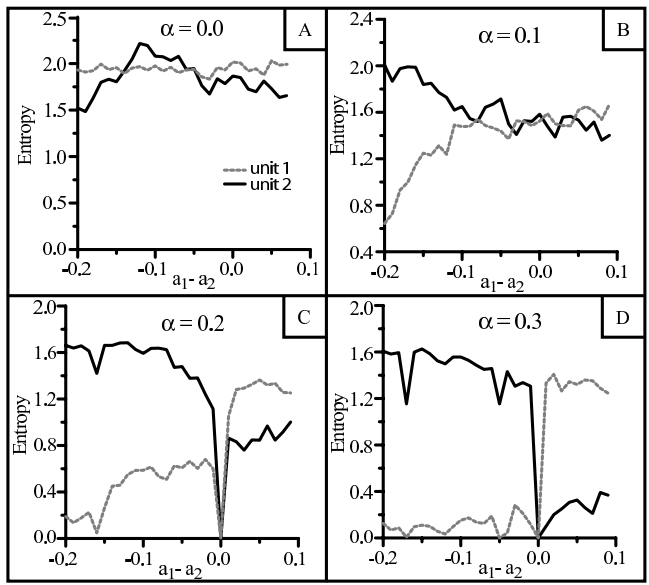}
\caption{The changes of the conditional entropy in the presence of coupling of different strengths. The control parameter $a_1$ was varied from 0.1 to 0.4, $a_2 = 0.3$ and was fixed. The coupling strength $\alpha$ is denoted on top of every panel.}
\end{figure}

We also investigated the effects of noise on the relative values of conditional entropies. A term of the form, $A * \psi_{1,2}(t)$ was added to the $y$-coordinate of each of the oscillators. The amplitude, $A$, varied from $0$ to $10$ and $\psi (t)\in[-1,1]$ was a randomly generated, uniformly distributed, variable (different for both oscillators). The observed phenomenon was very robust with respect to noise (Fig. 5). The noise effectively destroyed the synchronization at the point where the control parameters of the oscillators were equal (Fig. 5B-D). This is due to the fact that at this point the noise induced jitter in the timings of the events causing the same oscillator to sometimes precede and sometimes follow the other one. However, the salient features of the behavior remain the same: the unit with the higher excitation level always fires in a narrow window before the one with the lower excitation level and thus the timing interdependencies remain unchanged when the control parameters are different.  Thus, when the excitation levels of the oscillators are different, the conditional entropy of the unit with higher excitation is close to zero, whereas that of the other oscillator is high.    
\begin{figure}
\includegraphics[scale=.58]{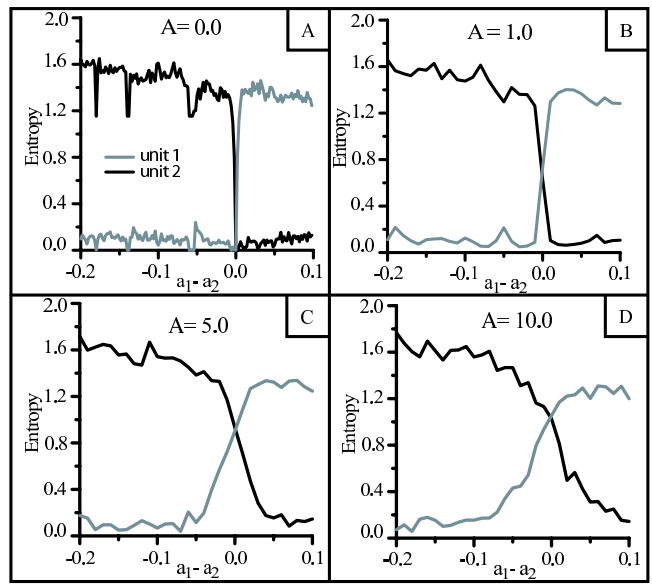}
\caption{Stability of the relative event timings in the presence of noise. The conditional entropy is measured for the varying relative excitation levels of R{\"o}ssler oscillators. The noise level is marked on each panel.}
\end{figure}

We show that the same phenomenon can be observed for a system of two interacting Hindmarsh-Rose models of the neuron. The equations for the coupled Hindmarsh-Rose neurons are:
\begin{equation}
\begin{array}{l}
\dot{x}_{1,2}= y_{1,2}-ax_{1,2}^3+bx_{1,2}^2-z_{1,2}+I_{01,2}+\alpha(x_{2,1}-x_{1,2})\\
\dot{y}_{1,2}= c-dx_{1,2}^2-y_{1,2}\\
\dot{z}_{1,2}= r\left[s(x_{1,2}-x_0)-z_{1,2}\right] 
\end{array}
\end{equation}
where the subscript denotes the neuron number; $a=1.0$, $b=3.0$, $c=1.0$, $d=5.0$, $r=0.006$, $s=4.0$, and $x_0=-1.6$ are the parameters of the model and $\alpha=1.1$ is the coupling strength. The parameter $I_0$ represents the amplitude of internal current to the neuron and is the control parameter of the system. We performed measurement of conditional entropy for two coupled neurons. Here, the event timings were the timings of spikes emitted by the neurons. The excitation levels of the two neurons varied (Fig 6). The conditional entropies changed in the same manner as in the case of the R{\"o}ssler oscillators. That is the conditional entropy of the neuron with the higher excitation level was zero and that of the other neuron is significantly higher (Fig. 6A, B). This suggests that the behavior is quite general and not limited to purely theoretical models. As in the case of the R{\"o}ssler oscillators, these dramatic changes in the CEs were due to the fact that the neuron with the higher excitation (i.e., higher $I_0$) fired a spike in a narrow window before the neuron with the lower excitation (Fig. 6C, D).  
\begin{figure}
\includegraphics[scale=.55]{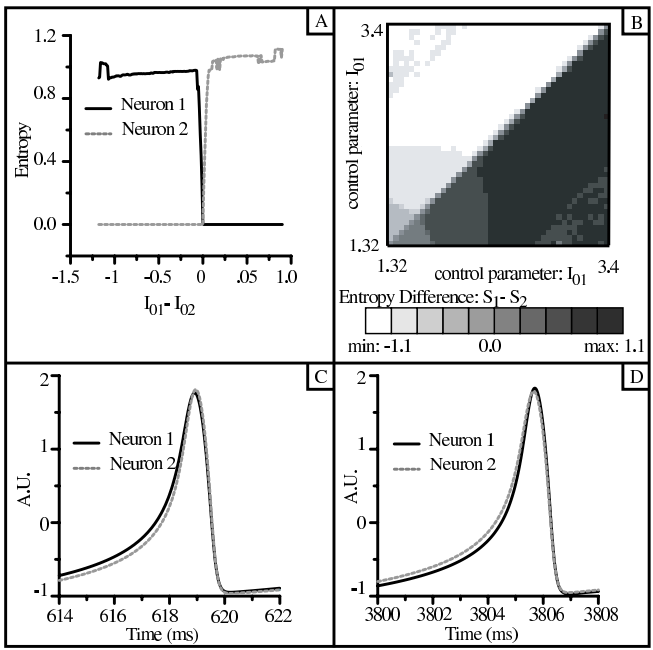}
\caption{Conditional entropy calculated for two Hindmarsh-Rose neurons with different relative excitation levels (value of $I_{01,2}$). Panel A: $I_{01}$ is varied from 1.32 to 3.4 and the control parameter of neuron 2 is fixed at $I_{02}=2.5$. Panel B: a grayscale map of the conditional entropy difference when both $I_{01,2}$ are varied from 1.32 to 3.4. Panels C and D: The relative timings of the spikes for the two neurons when the excitation is: C) $I_{01}>I_{02}$ and D) $I_{02}>I_{01}$.}
\end{figure}

These results imply that a link can be made between levels of excitation and the dramatic changes in the spike timing interdependencies of the coupled neurons. The small shifts observed and measured may have big implications in connection to experimentally observed activity dependent synaptic modification. Experimentally measured synaptic Long Term Potentiation (LTP) and Long Term Depression (LTD) as well as short term changes are directly linked to the relative timings of the emitted spikes \cite{poo, mcmah, abbott01}. If the presynaptic neuron emits a spike in a narrow window before the postsynaptic neuron the connecting synapse (and in some cases other synapses) will be potentiated. Conversely, if the postsynaptic neuron fires systematically before the presynaptic neuron the related synapses will be depressed. We, on the other hand show that relative spike timings depend upon the relative level of excitation between the neurons in question. Thus we established a possible link between the relative levels of excitation of the neurons themselves and the pattern of synaptic modifications. Furthermore, we can postulate that since the synaptic modifications will depend differentially on the relative excitation levels of the coupled neurons, the direction of information processing will be directly linked to them: neurons having higher excitation levels will effectively drive the neurons with lower excitation levels, since this monosynaptic pathway will be potentiated. Since it is known that those levels of excitation can be controlled intrinsically by neurons (through expression of appropriate receptors) \cite{turrigiano}, or externally through the signal arriving at the given neuron, this implies that the directionality of information processing can be dynamically controlled by the information itself. We implement this hypothesis on a model of two coupled neurons, with non-symmetrical modifiable couplings. Changes in the coupling strengths, $\alpha_{1,2}$, of both neurons are linked to their respective conditional entropies. The general form of those changes is given by:
$\alpha_{1,2}(S)=\frac{\lambda}{1+\beta S^2_{1,2}}$,
where $\lambda=1.1$ and $\beta=5.0$ are constants. Initially the neurons have the same coupling, but different control parameters, $I_{01}$ and $I_{02}$. The coupling $\alpha_1$ (driven by the value of conditional entropy) from the neuron with higher excitation level got progressively stronger whereas that of the other one became significantly smaller until a steady state was reached, where $\alpha_1$ achieved its maximal value and $\alpha_2$ decreased to its minimal one. At t=3000ms the control parameter of the neurons was switched, so that the neuron with the previously lower excitation level now had a higher one. The change evoked rapid synaptic modifications (changes in the coupling strength $\alpha_{1,2}$) driven by the changing conditional entropies. When steady state was reached again the synapses effectively switched their efficacy (the values of $\alpha_{1,2}$ had reversed). Thus the neuron that was initially driven became a driver (Fig. 7A, B) - the information flow is reversed. 
\begin{figure}
\includegraphics[scale=.55]{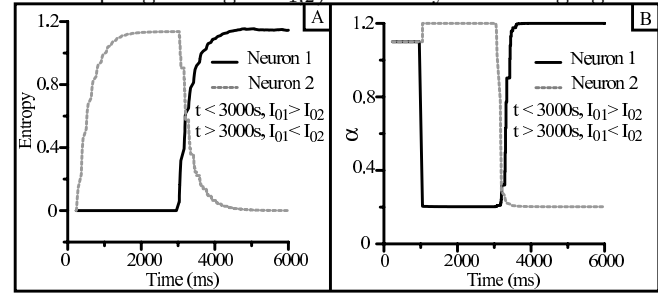}
\caption{Simulation of the modifiable non-symmetrical coupling between the neurons with changing excitation levels. The information flow reverses itself as the relative excitation levels are changed: A) conditional entropies and B) coupling strengths.}
\end{figure}

In conclusion, we have constructed a measure that monitors dynamically changing distributions of event timings of two coupled units. The use of this measure led to discovery of a relation between the event timings of two coupled units with different values of control parameters (i.e. excitation levels). The event timings of the unit with higher excitation level happen within a narrow window before those of the other unit. We have shown this phenomenon on two different systems: R{\"o}ssler oscillators and Hindmarsh-Rose models of the neurons. We believe that the observed phenomenon together with the known experimental results of activity dependent synaptic modification may provide an important link between dynamic modifications of the network during the information processing and the content of the processed information itself. The obtained results have to be now investigated experimentally.

\end{document}